\documentclass[10pt,journal]{IEEEtran}
\usepackage[T1]{fontenc}
\usepackage{cite}
\usepackage{amsmath,graphicx}
\usepackage{amssymb}
\usepackage{algpseudocode}
\usepackage{amsfonts}
\usepackage{graphicx}
\usepackage{fancyhdr}  %
\usepackage{cases}
\usepackage{extarrows}
\usepackage{algorithm}
\usepackage{multirow,tabularx}
\usepackage{mathtools}
\usepackage{xcolor}
\usepackage[english]{babel}
\usepackage{caption}
\usepackage{autobreak}
\usepackage{bm}

\newtheorem{theorem}{\bf Theorem}
\newtheorem{lemma}{\bf {Lemma}}
\newtheorem{proposition}{\bf {Proposition}}
\newtheorem{assumption}{\bf{Assumption}}
\newtheorem{definition}{\bf{Definition}}

\ifCLASSINFOpdf
\else
\fi
\begin{document}
\title{Stochastic Geometry Analysis for Distributed RISs-Assisted mmWave Communications}
\author{Yuan Xu\textsuperscript{1,2}, Li Wei\textsuperscript{3}, Chongwen Huang\textsuperscript{1,2}, Yongxu Zhu\textsuperscript{4}, Zhaohui Yang\textsuperscript{1}, Jun Yang\textsuperscript{5}, Jiguang He\textsuperscript{6}, 
\\Zhaoyang Zhang\textsuperscript{1} and M\'{e}rouane~Debbah\textsuperscript{7,8},~\IEEEmembership{Fellow,~IEEE}
\\\textsuperscript{1} College of Information Science and Electronic Engineering, Zhejiang University, 310027, Hangzhou, China
\\\textsuperscript{2} State Key Laboratory of Integrated Service Networks, Xidian University, 710071, Xi'an, China
\\\textsuperscript{3} School of Electrical and Electronics Engineering, Nanyang Technological University, 639798, Singapore
\\\textsuperscript{4} National Communications Research Laboratory, Southeast University, Nanjing 210096, China
\\\textsuperscript{5} Wireless Product R\&D Institute, ZTE Corporation, 518057, China
\\\textsuperscript{6} Technology Innovation Institute, 9639 Masdar City, Abu Dhabi, UAE
\\\textsuperscript{7} KU 6G Research Center, Khalifa University of Science and Technology, P O Box 127788, Abu Dhabi, UAE
\\\textsuperscript{8} CentraleSupelec, University Paris-Saclay, 91192 Gif-sur-Yvette, France
\vspace{-5mm}
\thanks{
The work was supported by the China National Key R\&D Program under Grant 2021YFA1000500 and 2023YFB2904800, National Natural Science Foundation of China under Grant 62331023, 62101492, 62394292, 62231009 and U20A20158, Zhejiang Provincial Natural Science Foundation of China under Grant LR22F010002, Zhejiang Provincial Science and Technology Plan Project under Grant 2024C01033, and Zhejiang University Global Partnership Fund.  
}
}
\maketitle
\thispagestyle{empty}
\pagestyle{empty}
\begin{abstract}
Millimeter wave (mmWave) has attracted considerable attention due to its wide bandwidth and high frequency. However, it is highly susceptible to blockages, resulting in significant degradation of the coverage and the sum rate. A promising approach is deploying distributed reconfigurable intelligent surfaces (RISs), which can establish extra communication links. In this paper, we investigate the impact of distributed RISs on the coverage probability and the sum rate in mmWave wireless communication systems. Specifically, we first introduce the system model, which includes the blockage, the RIS and the user distribution models, leveraging the Poisson point process. Then, we define the association criterion and derive the conditional coverage probabilities for the two cases of direct association and reflective association through RISs. Finally, we combine the two cases using Campbell's theorem and the total probability theorem to obtain the closed-form expressions for the ergodic coverage probability and the sum rate. Simulation results validate the effectiveness of the proposed analytical approach, demonstrating that the deployment of distributed RISs significantly improves the ergodic coverage probability by 45.4\% and the sum rate by over 1.5 times.
\end{abstract}


\vspace{-6mm}
\section{Introduction}\label{sec:intro}
\vspace{2.3mm}
Millimeter wave (mmWave) technology has become the core technology for the next generation of mobile communication networks due to its abundant spectrum resources and low spectrum interference\cite{Jameel8594703,Lodro8673447,Yang8613274}. The high-frequency nature of mmWave massive multiple-input multiple-output technology enables high spatial resolution and directivity, facilitating concentrated power transmission of communication signals. However, it renders the signals highly susceptible to atmospheric absorption and rain attenuation, resulting in significant penetration loss and path loss\cite{Kutty7342886,Niu07228}. To address these challenges, one feasible approach is to deploy distributed reconfigurable intelligent surfaces (RISs), each consisting of numerous low-cost passive reflective elements\cite{Huang8741198,9786794,9899454}. Using an intelligent controller, each unit or element can manipulate amplitudes and phase shifts to effectively change incident electromagnetic waves, thereby reconfiguring the wireless propagation environment. Moreover, the distributed RISs provide additional transmission links, effectively guaranteeing the quality of service for users\cite{9762646}. Therefore, we are interested in the performance gain that distributed RISs can bring to communication systems.
\par 
Nevertheless, analyzing the communication performance within a distributed RISs-assisted system remains a challenge. Recently, stochastic geometry has proven to be a successful mathematical framework for simulating the operation of diverse types of wireless networks\cite{Win2006ErrorPO,Win2009AMT, Haenggi5226957}. In \cite{bai6840343}, stochastic geometry was utilized to stochastically model the locations of receivers and transmitters using Poisson point processes (PPPs). By considering typical receivers positioned at the origin, the analysis of average network performance was simplified, provided performance metric bounds, and approached realistic scenarios. In \cite{Bai6932503}, the authors leveraged the line-of-sight (LoS) probability function proposed in \cite{bai6840343} to model the locations of LoS and non-LoS base stations (BSs) as two independent inhomogeneous PPPs, and further analyzed the dense network case by applying a simplified sphere model. In \cite{Rebato8628991}, the authors proposed an analytical framework that incorporates realistic channel models and antenna element radiation patterns. In \cite{Kishk2001}, the authors disregarded the small-scale channel fading model, deployed RISs as coatings on a portion of the blockages, and utilized stochastic geometry to analyze the impact of coating blockages with RISs on the cellular network coverage probability.  The authors in \cite{9110835} used stochastic geometry to analyze the performance of a realistic two-step user association-based mmWave network, consisting of multiple users, transmitters, and one-hop reflection from a large intelligent surface.
\par 
Building on the discussion above, this paper investigates the performance analysis in distributed RISs-assisted mmWave communication systems using stochastic geometry, and numerical results validate the gains from the deployment of distributed RISs. In more detail, we start with establishing the scene model and the system model using stochastic geometry. Next, we define the association criterion and decompose the RIS distribution model into several inhomogeneous PPPs using the LoS probability. Afterward, we analyze the two cases of direct association between the users with the serving BS and association through RIS reflection, and derive the association probabilities, the distance distributions, and the conditional coverage probabilities. Using the total probability theorem and Campbell's theorem, we combine the two cases to derive the ergodic coverage probability. Moreover, we derive closed-form expressions for the sum rate by transformations between the expectation and distribution functions. 
\par
Simulation results verify the gains in the ergodic coverage probability and the sum rate brought by the distributed RISs. Specifically, distributed RISs significantly improve the ergodic coverage probability by 45.4\% and the sum rate by over 1.5 times when the blockage density is $\lambda_b=1.59e$-$3/$m$^2$. These results indicate how many distributed RISs should be deployed to reach the performance limit, which provide a valuable guideline for the practical deployment of RISs in cellular networks.
\vspace{-3.5mm}
\section{System Model}\label{sec:format} 
\begin{figure}[t]
	\begin{center}
			\centerline{\includegraphics[width=0.45\textwidth]{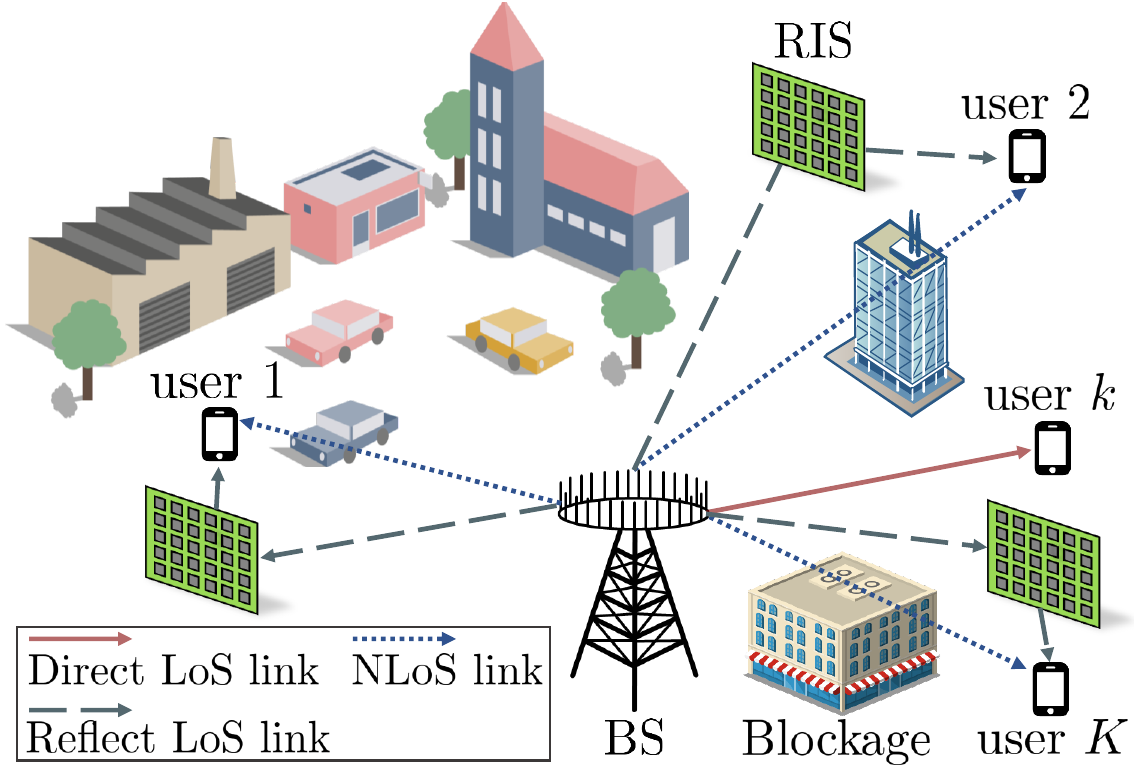}}  \vspace{-1mm}
			\caption{Downlink distributed RISs-assisted mmWave communications.  }
			\label{fig:scene} \vspace{-7mm}
		\end{center}
\end{figure}

\vspace{2.5mm} 
We consider a downlink distributed RISs-assisted mmWave communication system as illustrated in Fig.~\ref{fig:scene}, including the single BS serving multiple users. It is assumed that a BS can associate with several users, but only serves one user in a time slot, whether through a direct link or a RIS reflected link, providing the maximum channel gain. Considering the irregular shapes of real cells, we introduce the concept of the cell virtual radius, which is the radius of an equivalent circle of average cell size, denoted as $r_v$.
\par
Disregarding the height information, the cell is modeled as a planar area with a virtual radius of $r_v=R$. The BS is positioned at the center of this area, equipped with $N_{BS}$ antennas. Assuming users remain static within a short time interval, their positions are modeled by a PPP with a density of $\lambda_{u}(\xi)$, denoted as $\Phi_{u}=\{u_i\}\in\mathbb{R}^2$, where $u_i$ is the position coordinate of the $i$-th user, and $\Phi_{u}$ is the set that comprises all user coordinate points. The parameter $\xi$ denotes the distance between the BS and users, and each user is equipped with $N_u$ antennas. The positions of RISs are denoted as the PPP $\Phi_{R}=\{R_i\}\in\mathbb{R}^2$, with a density of $\lambda_R$. Here, $R_i$ is the position of the $i$-th RIS, $\Phi_{R}$ comprises all RIS coordinate points, and each RIS consists of $N_R$ elements. The horizontal and vertical spaces between these elements/antennas are $\lambda/2$, where $\lambda$ represents the signal wavelength. In addition, it is assumed that there is always a LoS link between the BS and the RISs, and that the BS can select different steering vectors as its transmit beams toward the RISs.
\par
Blockages, characterized by their spatial randomness and irregular shapes, especially for buildings in urban areas, are typically modeled as random object processes (ROPs) in the theory of random shapes\cite{Richard10}. However, analyzing ROPs can be complex, especially when considering correlations between objects, as well as their shape, position, and orientation. To make it analyzable, we approximate blockages using the line Boolean model\cite{bai6840343}, i.e., a line segment models each blockage.
\begin{assumption}
\textit{(Blockage Process)} The central coordinates of line segments are modeled by a PPP $\Phi_{b}=\{z_i\}\in\mathbb{R}^2 $ with a density of $\lambda_b $, where $z_i$ is the center coordinate of the $i$-th blockage, and $\Phi_b $ comprises all blockage center coordinates. The length of each blockage line segment, denoted as $L_i$, follows a uniform distribution $\mathcal{U}[L_{\text{min}}, L_{\text{max}}]$, where $L_{\text{min}}$ and $L_{\text{max}}$ represent the minimum and maximum lengths of all blockage line segments, respectively. The mean value of the length distribution is denoted as $E[L]$. The orientation of each blockage line segment is defined as the angle between the line segment and the positive direction of the $x$-axis. This orientation angle, denoted as $\phi_{b,i}$, is uniformly distributed and follows $\mathcal{U}[0,2\pi]$.
\end{assumption}
\begin{assumption}
	\textit{(Large-Scale and Small-Scale Fading Channel Gain) }\label{smallscale} The LoS channel gain for large-scale fading is described by the 3GPP model as $g=\frac{10^\alpha}{d^\beta}$. In this model, $\alpha$ and $\beta$ are parameters that depend on the specific environment and frequency band, and $d$ represents the distance between the transmitter and receiver. For small-scale fading, the channel fading gain is assumed to follow an exponential distribution\cite[Remark 5]{Rebato8628991} with a mean value of $\rho_{t,r}$, which represents the directional gain. To express this directional gain, the practical array pattern is approximated by the segmented antenna model, we define $M_t,m_t,\psi_t\ (M_r,m_r,\psi_r)$ as the main lobe gain, side lobe gain, and main lobe beam width at the transmitter (receiver), respectively. For a beam-aligned link, the directional gain is expressed as $\rho_{t,r}=M_tM_r$. Meanwhile, the angle of arrival (AOA) and the angle of departure (AOD) of a non-beam-aligned link follow a uniform distribution on the interval $(0,2\pi)$, and its directional gain is expressed as\vspace{-0.5mm}
	\begin{equation}
		\rho_{t,r}=\left\{
		\begin{aligned}
			&M_tM_r,\ \text{w.p.}\  \frac{\psi_t}{2\pi}\frac{\psi_r}{2\pi}\\
			&M_tm_r,\ \text{w.p.}\  \frac{\psi_t}{2\pi}(1-\frac{\psi_r}{2\pi})\\
			&m_tM_r,\ \text{w.p.}\  (1-\frac{\psi_t}{2\pi})\frac{\psi_r}{2\pi}\\
			&m_tm_r,\ \text{w.p.}\  (1-\frac{\psi_t}{2\pi})(1-\frac{\psi_r}{2\pi}).
		\end{aligned}
		\right.
	\end{equation}
 \par
	For an isotropic antenna, the main lobe gain $M_t$ and $M_r$ are equal to the number of antennas $N_t$, $N_r$ at the transmitter and receiver, and the side lobe gain is $m_t=1/\sin^2(\frac{3\pi}{2\sqrt{N_t}}) $, $m_r=1/\sin^2(\frac{3\pi}{2\sqrt{N_r}}) $.
\end{assumption}

\par
As for the signal model, it is assumed that the data transmission links between the BS and the users have already been beam-aligned using existing beamforming techniques like zero forcing\cite{Spencer1261332}, minimum mean square error\cite{Nguyen8048606}, single value decomposition\cite{Li7996970}, etc. However, interfering signals may arrive through non-beam-aligned links.
\par
Thus, the signal-to-interference-plus-noise ratio (SINR) of the direct links and the RIS reflection links can be defined as follows. For a user located at a distance $\xi$ from the BS, the large-scale fading channel gain of the direct link is denoted as $g_d=\frac{10^\alpha}{\xi^\beta}$. The received signal at the user is denoted as $y_d$ with power $|y_d|^2=\frac{10^\alpha\cdot P_0\cdot h_d}{\xi^\beta}$, where $P_0$ is the transmit power of the BS to each user and $h_d\sim \mathbf{Exp}(N_{BS}N_u)$ is the small-scale fading channel gain of the BS-user link.
It is assumed that the interference from other users has been eliminated using existing techniques. Therefore, the SINR of the direct link is expressed as
\begin{equation}
    \gamma_d=\frac{10^\alpha\cdot P_0\cdot h_d}{\xi^\beta\sigma^2} ,
\end{equation}
where $\sigma^2$ is the noise power.
\par
For the case where the BS reflects the signal to users through the $l$-th RIS $R_l$, we define the channel as the $l$-th reflected LoS link. The large-scale fading channel gain and received signal power of the $l$-th reflected link are expressed as $g_{I_l}=\frac{10^{2\alpha}}{(s_lr_l)^\beta}$ and $|y_{I_l}|^2=\frac{10^{2\alpha}P_0 h_{s_l} h_{r_l}}{(s_lr_l)^\beta} $, respectively. Here, $s_l$ and $h_{s_l}\sim \mathbf{Exp}(N_{BS}N_R)$ are the distance and the small-scale fading channel gain of the BS-$R_l$ link, $r_l$ and $h_{r_l}\sim \mathbf{Exp}(N_RN_u)$ are the distance and the small-scale fading channel gain of the $R_l$-user link, respectively. The SINR of the BS-$R_l$-user link can be expressed as 
\begin{equation}
    \gamma_{I_l}=\frac{10^{2\alpha}P_0 h_{s_l} h_{r_l}}{(s_lr_l)^\beta\sigma^2}.
\end{equation}
\vspace{-4.7mm}
\section{Performance Analysis}
\vspace{2mm}
In this section, a theoretical performance analysis is presented leveraging the inherent void probability and contact distribution\cite{MOLTCHANOV20121146} of the PPP, the location-dependent thinning\cite{RePEc}, and the LoS probability.
\begin{proposition}\label{thinning}
	\textit{(Location-Dependent Thinning}\cite{RePEc}\textit{)} For a homogeneous PPP $\Phi$ with density $\lambda$, thinning with probability $g(x)$ that depends only on point $x$ and not on the rest of the points, results in an inhomogeneous PPP $\tilde{\Phi}$, with density $\tilde{\lambda}=[1-g(x)]\lambda$. Simultaneously, the removed points constitute an inhomogeneous PPP with density $\overline{\lambda}=g(x)\lambda$.
\end{proposition}

\begin{assumption}
	\textit{(LoS Probability) \label{PLoS}}For each link, the LoS probability is only related to the link distance $d$ and blockage information, namely, $\lambda_b$, $E[L]$, but not to the channel state information. Furthermore, the influence of the correlation of the blockages between different links can be ignored. Let $Z$ denote the number of blockages intersecting the link. According to the line Boolean model of blockages, we have $E[Z]=\frac{2\lambda_b E[L]d}{\pi}$\cite{bai6840343}. Consequently, the probability of a LoS link existing at a distance $d$ is equivalent to the probability of no blockage intersecting that link, denoted as the void probability when $Z=0$, which is represented as $P_{LoS}(d)=\exp(-E[Z])=\exp(-\frac{2\lambda_b E[L]d}{\pi})$.
\end{assumption}
\par
\begin{figure}[t]
    \begin{center}
        \centerline{\includegraphics[width=0.45\textwidth]{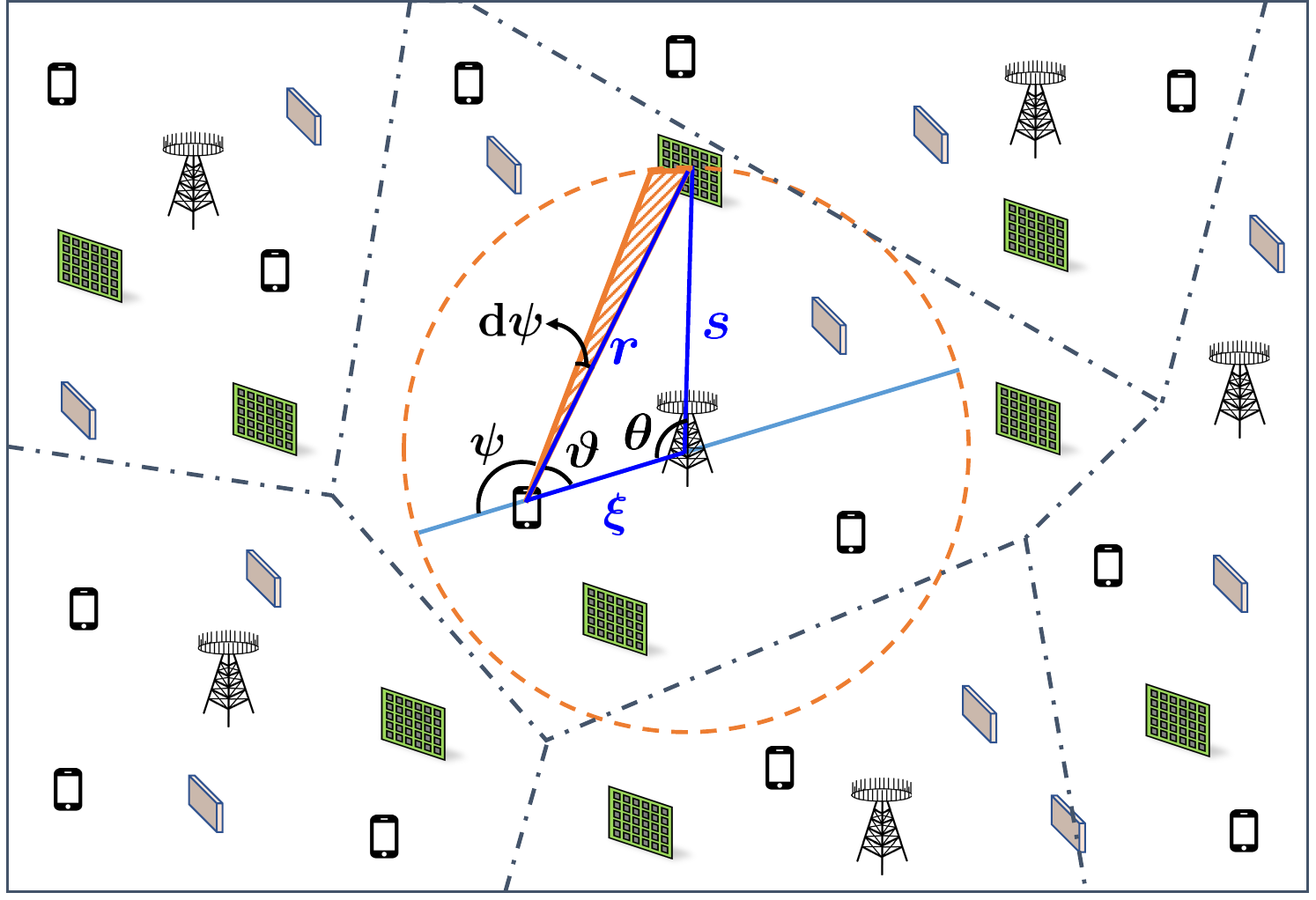}}  \vspace{-1mm}
        \caption{The cell-wide integration of ${\lambda}^L_R$. }
        \label{fig:P_R_s} 
    \end{center}\vspace{-7mm}
\end{figure}

In the following, we start with deriving the probability that RISs can assist the communication for any user within the set $\Phi_u$. It is assumed that RISs can both reflect and transmit to achieve 360-degree coverage. It is further assumed that RISs only assist communication in scenarios where the direct link is non-line-of-sight (NLoS). 
\begin{lemma}
	\textit{(RISs Assist Probability)} In the depicted scenario in Fig.~\ref{fig:P_R_s}, the distances of the BS-user, BS-RIS, and RIS-user links are $\xi$, $s$, and $r$, respectively. We introduce the angle between the BS-user link and the RIS-user link as $\vartheta$, and define $\psi=2\pi - \vartheta$. Since we assume that there is always a LoS link between the BS and RISs, the probability that the RIS can provide a reflection link for the BS and the user is $P_{LoS}(r)$. Employing location-dependent thinning, which preserves the RIS PPP $\Phi_R$ with the probability $P_{LoS}(r)$, we obtain an inhomogeneous PPP $\Phi_R^L$ with a density of $\lambda_R^L=\lambda_R\cdot P_{LoS}(r) $. RISs in $\Phi_R^L$ can assist the communication between the BS and the user.
\end{lemma}
\par
Next, we define the association criterion.
\begin{definition}
	\textit{(Association Criterion)} \label{asscociaion criterion}Each user communicates with its serving BS through the link providing the highest signal power, either a direct link or a reflected link,\vspace{-1mm}
	\begin{equation}
		|y^*|^2=\max(|y_{d}|^2,\max\limits_{R_l\in\Phi_R^L} |y_{I_{l}}|^2).
	\end{equation}
 \par\vspace{-1mm}
	To formulate the feasible analysis equations, we introduce suitable simplifications by substituting the small-scale fading channel gain with its time-averaged value. Therefore, the association criterion can be simplified as \vspace{-1mm}
 \begin{equation}
 \begin{split}
     |y^*|^2&=\max(g_{d}N_{BS}N_u,\max\limits_{R_l\in\Phi_R^L} g_{I_{l}}N_{BS}N_RN_RN_u)\\
     &=N_{BS}N_u\max(g_{d},\max\limits_{R_l\in\Phi_R^L} g_{I_{l}}N_R^2).
 \end{split}
 \end{equation}
\end{definition}
\begin{lemma} \textit{(Reflection Probability)} \label{reflection probability}
	For a user at a distance $\xi$ from the BS, the probability of existing at least one RIS reflective LoS link within the cell is 
	\begin{equation}
         \begin{split}
             P_R^{s}(\xi)\!=&1\!-\exp(-\!\!\int_0^{2\pi}\!\!\!\!\!{\lambda}^L_R\frac{\!(\!\sqrt{R^2\!-\!\xi^2\sin^2\!\psi}\!-\!\xi\cos\!\psi)^2\!\!}{2}\mathrm{d}\psi).
         \end{split}
	\end{equation}\vspace{-3mm}
\end{lemma}
\par
\textit{Proof:} First, we calculate the number of RISs capable of providing reflective LoS links, which requires a cell-wide integration over the density of LoS RISs ${\lambda}^L_R$. Since ${\lambda}^L_R$ depends on the RIS-user distance $r$, which requires transforming the area into an expression on $r$. To facilitate this transformation, we introduce the angle $\psi=2\pi-\vartheta$, and the shaded area in Fig.~\ref{fig:P_R_s} can be approximated as $\frac{r^2\mathrm{d}\psi}{2} $. Utilizing the law of cosines, we can obtain $\cos(2\pi-\psi)=\cos\vartheta=\frac{\xi^2+r^2-R^2}{2\xi r} $. Further, we derive that $r=\sqrt{R^2-\xi^2\sin^2 \psi}-\xi\cos\psi $. Hence, the number of RISs that can provide reflective LoS links is
\begin{equation}
	\begin{split}
		L^{s}(\xi)=&\int_0^{2\pi} {\lambda}^L_R\frac{r^2\mathrm{d}\psi}{2}\\
		=&\int_0^{2\pi} {\lambda}^L_R\frac{(\sqrt{R^2-\xi^2\sin^2 \psi}-\xi\cos\psi)^2}{2}\mathrm{d}\psi.
	\end{split}
\end{equation}
\par
Thus, the probability is further expressed as
\begin{equation}
\centering
	\begin{split}
		&P_R^{s}(\xi)=1-\exp(-L^{s}(\xi))\\
		&=1\!-\exp(-\int_0^{2\pi}\!\! {\lambda}^L_R\frac{(\sqrt{R^2-\xi^2\sin^2 \psi}-\xi\cos\psi)^2\!\! }{2}\mathrm{d}\psi).
	\end{split}
\end{equation}
\begin{lemma}
	\textit{(Distance Distribution of Reflected LoS Links)} \label{reflective LoS link dist} Given the distance between the BS and a user as $\xi$, the distance between the BS and the $l$-th RIS as $s_l$, and the distance between the user and the $l$-th RIS as $r_l$, we define $\eta\triangleq\min\limits_l\{s_l\cdot r_l\}$, its cumulative distribution function is derived in Eq.~(\ref{eq:F_eta}), where $\cos(\theta)\in[\frac{s^2+\xi^2}{2s\xi},\frac{s^4+s^2\xi^2-x^2}{2s^3\xi}]$.
 \begin{figure*}[t]\vspace{-2mm}
     \centering
     \begin{equation}\label{eq:F_eta}
		F_{\eta|\xi}(x)=\left\{
		\begin{aligned}
			0 ,& \quad x\leq r\\
			1-\exp(-\int_0^R\int_{\theta}\lambda_RP_{LoS}(\sqrt{s^2+\xi^2-2s\xi\cos\theta})s\mathrm{d}\theta\mathrm{d}s) ,& \quad \text{else}\\
		\end{aligned}
		\right.
	\end{equation}
 \begin{equation}\label{eq:proof_F_eta}
	\begin{aligned}
		&F_{\eta|\xi}(x)=Pr(\min\limits_l\{s_l\cdot r_l\}\leq x|\xi)=1-Pr(\min\limits_l\{s_l\cdot r_l\}> x|\xi)\\
		&\overset{(a)}{=}1-Pr_{void}(\frac{s^4+s^2\xi^2-x^2}{2s^3\xi} < \cos\theta < \frac{s^2+\xi^2}{2s\xi})\\
		&=\left\{
		\begin{aligned}
			0 ,& \quad x\leq \xi\\
			1-\exp(-\int_0^R\int_\theta{\lambda}^L_R s\mathrm{d}\theta \mathrm{d}s) ,& \quad \text{else}\\
		\end{aligned}
		\right.\\
		&=\left\{
		\begin{aligned}
			0 ,& \quad x\leq \xi\\
			1-\exp(-\int_0^R\int_\theta\lambda_RP_{LoS}(\sqrt{s^2+\xi^2-2s\xi \cos\theta})s\mathrm{d}\theta\mathrm{d}s) ,& \quad \text{else}\\
		\end{aligned}
		\right.
	\end{aligned}
\end{equation}
 \end{figure*}
\end{lemma}
\par
\textit{Proof:}	As shown in Fig.~\ref{fig:P_R_s}, the angle between the BS-RIS and BS-user links is defined as $\theta$, because the location-dependent thinning is only related to the distance but not to the angle, it can be assumed that $\cos\theta\sim\mathcal{U}[-1,1]$, then we have $r_l=\sqrt{s_l^2+\xi_l^2-2s_l\xi_l\cos\theta}$. In Eq.~(\ref{eq:proof_F_eta}), the lower bound of equal sign $(a)$ is established because $\min\limits_l\{s_l\cdot r_l\}>x$. This implies that for $\forall l\in\{1,...R\}$, there is no $s_l\cdot r_l<x$. Omitting the subscript $l$, $s\cdot r<x$ is equivalent to $\cos\theta > \frac{s^4+s^2\xi^2-x^2}{2s^3\xi}$. The upper bound is derived from the fact that $r^2=s^2+\xi^2-2s\xi\cos\theta>0$. Thus, within the region satisfying $\frac{s^4+s^2\xi^2-x^2}{2s^3\xi} < \cos\theta < \frac{s^2+\xi^2}{2s\xi}$ and in a radius $R$, there exists no RIS capable of assisting communication. Consequently, the number of points in ${\Phi}^L_R$ is 0.

\par
\begin{definition}
	Define the coverage probability as 
	\begin{equation}
		P_{cov}(T)=Pr(\gamma>T),
	\end{equation}
where $T$ denotes the set threshold, and $\gamma$ denotes the SINR of the associated link, which could be either a direct link or a reflected link.
\end{definition}
\begin{theorem}
	\textit{(Ergodic Coverage Probability)}\label{eq:coverage single}
    The ergodic coverage probability of the cell is
	\begin{equation}
		E[P_{cov}^s(T)]=\frac{\int_0^R P_{cov|\xi}^s(T)\lambda_u(\xi)2\pi \xi\mathrm{d}\xi}{\int_0^R \lambda_u(\xi)2\pi \xi\mathrm{d}\xi},
	\end{equation}
	where $P_{cov|\xi}^s(T)$ is the coverage probability of users at distance $\xi$ from the BS, as expressed in Eq.~(\ref{eq:P_cov_xi}), and $\tau_1\triangleq\frac{\xi^\beta\sigma^2T}{P_0 10^\alpha}$, $\tau_2\triangleq(\frac{P_0 h_s h_r 10^{2\alpha}}{\sigma^2T})^\frac{1}{\beta} $.
    \begin{figure*}\vspace{-6mm}
        \centering
        \begin{subequations}\label{eq:P_cov_xi}
            \begin{equation}
		\begin{split}
			&P_{cov|\xi}^s(T)=P_{LoS}(\xi)\exp(-\frac{\tau_1}{N_{BS}N_u})+(1-P_{LoS}(\xi))P_R^s(\xi)\int_0^\infty\int_0^\infty F_{\eta|\xi}(\tau_2)f_{h_s}(x_1)\mathrm{d}x_1f_{h_r}(x_2)\mathrm{d}x_2
		\end{split}
	\end{equation}
	\begin{equation}
		f_{h_s}(x_1)=\left\{
		\begin{aligned}
			\frac{1}{N_{BS}N_R}\exp(-\frac{1}{N_{BS}N_R} x_1),&\ x_1>0\\
			0,&\ \text{else}
		\end{aligned}
		\right.
	\end{equation}
	\begin{equation}
		f_{h_r}(x_2)=\left\{
		\begin{aligned}
			\frac{1}{N_RN_{u}}\exp(-\frac{1}{N_RN_{u}} x_2),&\ x_2>0\\
			0,&\ \text{else}
		\end{aligned}
		\right. \vspace{-3mm}
	\end{equation}
        \end{subequations}
        {\noindent} \rule[-10pt]{17.5cm}{0.05em}\vspace{-5mm}\\
    \end{figure*}
    	
\end{theorem}
\par
\textit{Proof:} Since the multiplicative path loss of the RIS reflection link is several orders of magnitude larger than that of the direct link, the association criterion degenerates. When a direct LoS link between the user and the serving BS exists, there is a direct association. On the other hand, when the direct link is NLoS and a LoS reflection link exists, the association criterion is an association through RIS reflection.
\begin{lemma}
    Given that the distance between a user and its serving BS is $\xi$, the direct association probability is 
    \begin{equation}
        P_{Ad}(\xi)=P_{LoS}(\xi).
    \end{equation}
\end{lemma}
\par
Then, the conditional coverage probability of the direct link is derived as \vspace{-1mm}
\begin{subequations}
    \begin{equation}
\begin{split}
    &P_{cov_d|\xi}^s=Pr(\gamma_d>T)\\
    &=Pr(h_d>\frac{\xi^\beta\sigma^2T}{P_0 10^\alpha})=1-F_{h_d}(\frac{\xi^\beta\sigma^2T}{P_0 10^\alpha}) ,
\end{split}
\end{equation}
\begin{equation}
	F_{h_d}(x_0)=\left\{
	\begin{aligned}
		1-\exp(-\frac{1}{N_{BS}N_u} x_0),&\ x_0>0\\
		0,&\ \text{else} .
	\end{aligned}
	\right.
\end{equation}
\end{subequations}
\par
\begin{lemma}
    Given that the distance between a user and its serving BS is $\xi$, the probability that the serving BS and the user are associated through the reflected LoS link is 
    \begin{equation}
        P_{AI}(\xi)=(1-P_{LoS}(\xi))P_R^s(\xi),
    \end{equation}
where $(1-P_{LoS}(\xi))$ represents the probability of the direct link between the BS and the user is NLoS. In this scenario, the probability that there exists a reflected LoS link between the user and the NLoS BS is denoted as $P_R^s(\xi)$.
\end{lemma}
\par
Since the criterion prioritizes the link with the largest received signal power, we obtain the associated RIS as 
\begin{equation}
    l^*=\arg\max\limits_{l\in\{1,...L\}} \big\{\frac{10^{2\alpha}}{(s_lr_l)^\beta}N_R^2\big\}=\arg\min\limits_{l\in\{1,...L\}} \{s_lr_l\}.
\end{equation}
\par
The SINR of the associated link is expressed as $\gamma_I=\frac{10^{2\alpha}P_0  h_{s_{l^*}} h_{r_{l^*}}}{(s_{l^*}r_{l^*})^\beta\sigma^2}$. Thus, the conditional coverage probability of the reflected LoS link is
\begin{equation}
	\begin{split}
		P_{cov_I|\xi}^s=&Pr(\gamma_I>T)=Pr(\eta<(\frac{P_0 h_s h_r 10^{2\alpha}}{\sigma^2T})^\frac{1}{\beta})\\ 
		=&\int_0^\infty\int_0^\infty F_{\eta|\xi}(\tau_2)f_{h_s}(x_1)\mathrm{d}x_1f_{h_r}(x_2)\mathrm{d}x_2.
	\end{split}\nonumber
\end{equation} 
\par
Therefore, the conditional coverage probability of users at a distance $\xi$ from the BS is
\begin{equation}\vspace{2mm}
\begin{split}
    &P_{cov|\xi}^s(T)\\
    =&P_{Ad}(\xi)P_{cov_d|\xi}^s(T)+P_{AI}(\xi) P_{cov_I|\xi}^s(T)\\
    =&P_{LoS}(\xi)P_{cov_d|\xi}^s(T)+(1-P_{LoS}(\xi))P_R^s(\xi) P_{cov_I|\xi}^s(T).\nonumber
\end{split}
\end{equation} 
\par
The ergodic coverage probability of the cell is 
\begin{equation}
	E[P_{cov}^s(T)]\!=\!\frac{E[\sum\limits_{u_i\in\Phi_u}P_{cov|\xi}^s(T)]}{\int_0^R \lambda_u(\xi)2\pi \xi\mathrm{d}\xi}\!\!\overset{(b)}{=}\!\!\frac{\int_0^R P_{cov|\xi}^s(T)\lambda_u(\xi)2\pi \xi \mathrm{d}\xi}{\int_0^R \lambda_u(\xi)2\pi \xi\mathrm{d}\xi},\nonumber
\end{equation}
where $(b)$ is based on Campbell's theorem\cite{keeler2015campbell}.
\par
\begin{definition}
    The ergodic rate is defined as 
\begin{equation}
	R=E[W\log_2(1+\gamma)],
\end{equation}
where $W$ is the bandwidth used by each user.
\end{definition} 

\begin{proposition}
	For a positive random variable $X$,\label{positive RV}\vspace{-1mm}
        \begin{equation}
        E[X]=\int_0^\infty Pr(X>t)\mathrm{d}t . 
        \end{equation}
\end{proposition}
\begin{theorem}
	The sum rate of the cell is\label{eq:sum-rate single}
	\begin{equation}
		{\Upsilon}=\int_0^R \int_0^\infty P_{cov|\xi}^s(2^{\frac{t}{W}}-1)\mathrm{d}t \lambda_u(\xi) 2\pi \xi\mathrm{d}\xi .
	\end{equation}
\end{theorem}
\par
\textit{Proof:}	Obviously, the random variable $W\log_2(1+\gamma) $ is positive. According to Proposition \ref{positive RV}, the ergodic rate of users at a distance $\xi$ from the BS is\vspace{-1mm}
\begin{equation}
	\begin{split}
		R(\xi)&=E[W\log_2(1+\gamma)]\\
            &=\int_0^\infty Pr(W\log_2(1+\gamma)>t)\mathrm{d}t\\
		&=\int_0^\infty Pr(\gamma>2^{\frac{t}{W}}-1)\mathrm{d}t\\
		&=\int_0^\infty P_{cov|\xi}^s(2^{\frac{t}{W}}-1)\mathrm{d}t .
	\end{split}\nonumber
\end{equation}
\par
\vspace{-0mm}
Since the distance $\xi$ between each user and the BS is also a random variable, the sum rate of the cell can be expressed as\vspace{-2mm}
\begin{equation}
	\begin{split}
		{\Upsilon}=&\sum\limits_{u_i\in\Phi_u} E[R(\xi)]=E[\sum\limits_{u_i\in\Phi_u} R(\xi)]\\
            =&\int_0^R R(\xi)\lambda_u(\xi)2\pi \xi\mathrm{d}\xi\\
		=&\int_0^R \int_0^\infty P_{cov|\xi}^s(2^{\frac{t}{W}}-1)\mathrm{d}t \lambda_u(\xi)2\pi \xi\mathrm{d}\xi.
	\end{split}\nonumber
\end{equation}

\vspace{-4mm}
\section{Simulation Results}\label{sec:simulation_result}
\vspace{2mm}
In this section, we provide numerical results for the coverage probability and the sum rate in distributed RISs-assisted mmWave wireless communication systems. We assume that the mmWave cellular network operates at a frequency of $f_{c}=3$ GHz. The parameters in the large-scale fading channel gain model are set to $\alpha = 3.8$ and $\beta = 2.2$. The average length of the blockage line Boolean model is $E[L] = 15$ m. The BS antenna array consists of $N_{BS} = 64$ antennas. We set the virtual radius as $R = 100$ m and the density of users as $\lambda_u=3.18e$-$3/$m$^2$. Each user is equipped with $N_u = 4$ antennas and is allocated a bandwidth of $W = 200$ MHz.
\par
\begin{figure}[t]
	\centering
	\vspace{-4mm}
	\includegraphics[width=0.4\textwidth]{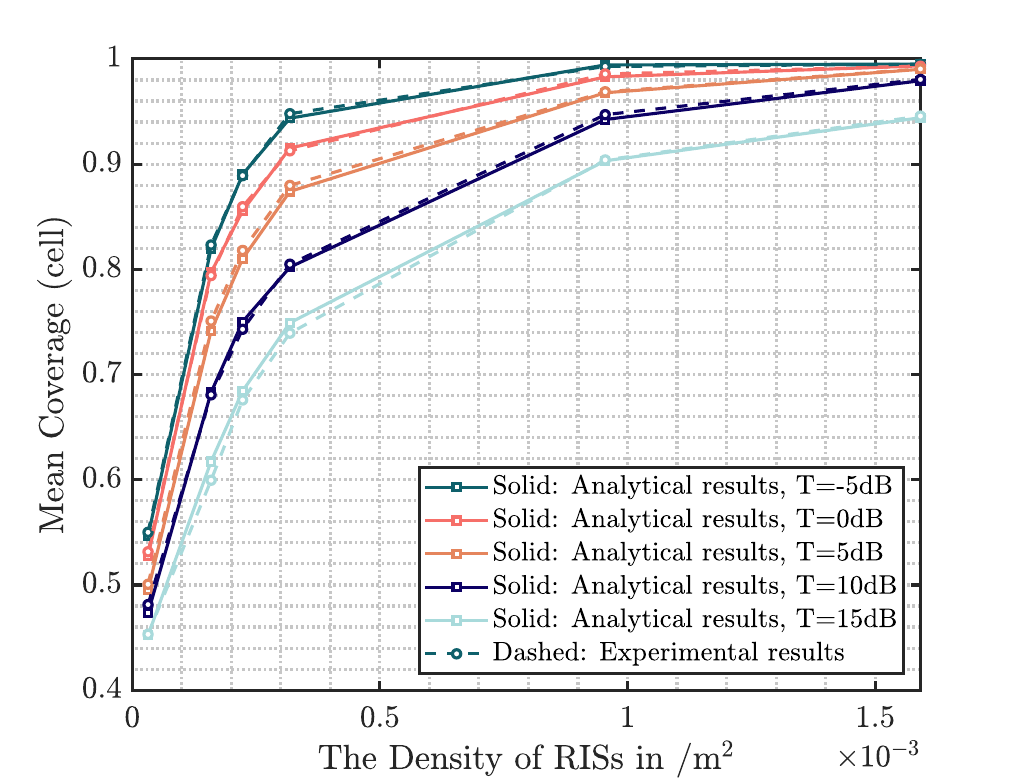}\vspace{-1mm}
	\caption{Effects of the density of RISs on the ergodic coverage probability of the cell.}\label{fig:cov_s}
\end{figure}

Fig.~\ref{fig:cov_s} plots the ergodic coverage probability for different densities of RISs with a blocking density of $\lambda_b=1.59e$-$3/$m$^2$. The solid line represents the ergodic coverage probability in Theorem~\ref{eq:coverage single}, while the corresponding dashed line is obtained by Monte Carlo simulation. The simulation results can overlap the theorem well, which proves the accuracy of our derived formulation. Moreover, we observe that deploying more RISs increases the coverage probability, but the rate of improvement decreases as the density of RISs increases, eventually reaching saturation. Specifically, deploying RISs with a density of $\lambda_R=1.59e$-$4/$m$^2$ (around the number of $5$) can improve the coverage probability by 27.3\%, while deploying RISs with a density of $\lambda_R=9.55e$-$4/$m$^2$ (around $30$) improves the coverage probability by 45.4\%.
\par 
\begin{figure}[t]\vspace{-5mm}
	\begin{center}
		\centerline{\includegraphics[width=0.4\textwidth]{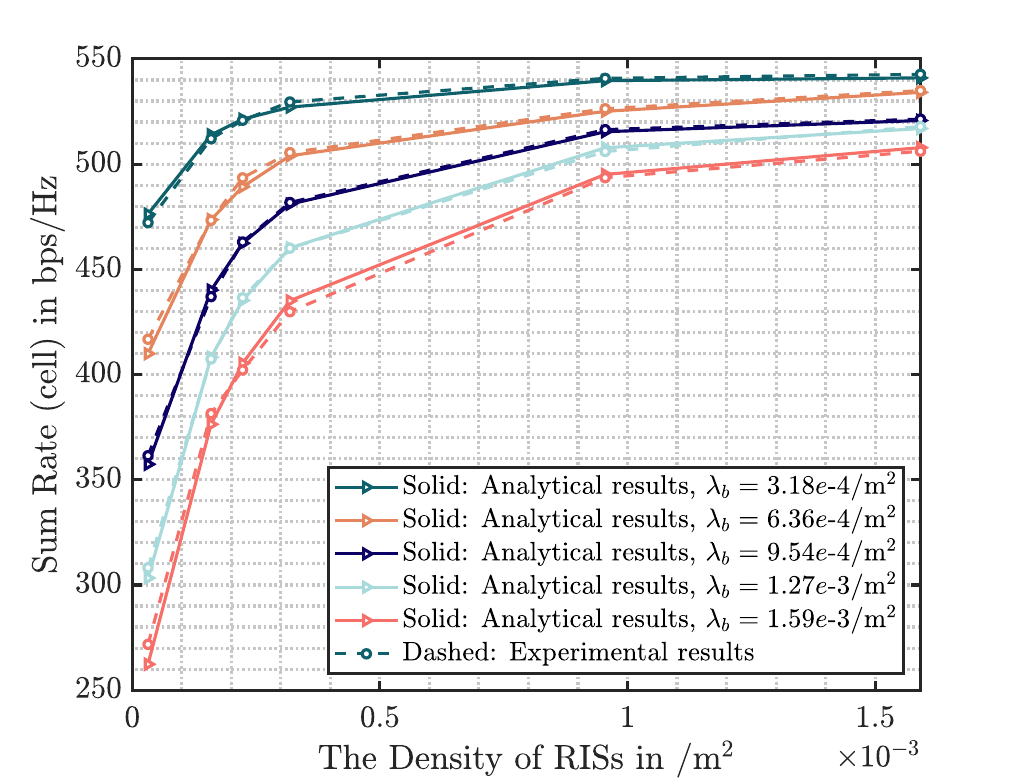}}  \vspace{-1mm}
		\caption{Effects of the density of RISs on the sum rate for different blockage densities.}
		\label{fig:rate_s} 
	\end{center}\vspace{-8mm}
\end{figure}
Fig.~\ref{fig:rate_s} plots the effect of the density of deployed RISs on the sum rate of all users in the cell for varying blockage densities. The results of Theorem \ref{eq:sum-rate single}, denoted by the solid line, and the Monte Carlo scenario simulation, denoted by the dashed line, are closely matched. As the blockages become denser, the sum rate decreases, yet the improvement in the sum rate is more significant with the same density of RISs deployed. For example, when RISs with density $\lambda_R=1.59e$-$3/$m$^2$ (around $50$) are deployed, the sum rate increases by $70.5$ bps/Hz (1.15 times) when $\lambda_b = 3.18e$-$4/$m$^2$, and by $241.8$ bps/Hz (1.91 times) when the blockage density $\lambda_b=1.59e$-$3/$m$^2$. However, it is worth noting that the curve tends to flatten out as the density of RISs increases, which means that the sum rate does not increase indefinitely as the density of RISs increases, but rather reaches an upper limit.

\vspace{-1.4mm}
\section{Conclusion}\label{sec:conclusion}
\vspace{1.4mm}
In this paper, we leveraged stochastic geometry to investigate the ergodic coverage probability and the sum rate in distributed RISs-assisted wireless communication systems. Firstly, the system model was established stochastically, including the distribution models of blockages, RISs, and users. Subsequently, the association criterion was defined, and the inhomogeneous PPPs for RISs were derived based on the LoS probability. Then, the association probabilities, the distance distributions, and the conditional coverage probabilities were obtained for the two cases of direct association and reflective association by RISs. Finally, we combined the two cases and obtained the closed-form expressions for the ergodic coverage probability and the sum rate. 
\par 
The stochastic geometry analysis and simulation results can provide insights into how many RISs should be distributed to achieve a given performance at different blockage densities, which essentially facilitates the optimal design and cost control for practical RIS deployment.

\bibliography{IEEEabrv,myrefs}
\bibliographystyle{IEEEtran}
\end{document}